Physical Properties of Sulfur Near the Polymerization Transition.

V. F. Kozhevnikov[1], W. B. Payne[1], J.K. Olson[1], C.L. McDonald[2], and C.E. Inglefield[2]

[1] University of Utah, Department of Physics, Salt Lake City, UT 84112
[2] Weber State University, Department of Physics, Ogden, UT 84408

Acoustical measurements, electron spin resonance, and Raman spectroscopy have been employed to probe sulfur over the temperature range 80° to 180° C, which includes the polymerization transition and the supercooled liquid state. Acoustical properties (sound velocity, absorption and impedance) have been studied with both longitudinal and transverse waves at frequencies between 500 kHz and 22 MHz. The results confirm that polymeric sulfur is a solution of long chain molecules in monomeric solvent, and that the polymerization transition is not a second order phase transition, as was proposed theoretically. Sulfur is a viscous liquid, but not viscoelastic, both below and above the polymerization transition temperature. It is shown that the classical Navier-Stokes theory is not applicable to the sound absorption in liquid sulfur in the highly viscous state.

**INTRODUCTION**

Sulfur, one of the most interesting elements of the periodical table, exhibits a wide variety of allotropes of cyclic molecules ($S_x$, with $x$ from 6 to 35) and the unusual phenomenon of equilibrium polymerization in the liquid state [1,2]. There is additional interest in understanding this transition because equilibrium polymerization also occurs in organic and biological systems [3]. Equilibrium polymerization can be viewed as a particular example of the general tendency in condensed matter for particles to cluster [4]. Sulfur is the simplest substance exhibiting this phenomenon, and therefore should be the most convenient to model. In addition, the polymerization transition in sulfur is an example of a liquid-liquid phase transition. Such transitions are currently of considerable interest [5].

Polymerization in sulfur has been studied for many years (see Refs 1-3 and references therein), but a coherent picture of this phenomenon is still missing. Sulfur melts at about 120° C, forming a light-yellow low-viscosity liquid mostly consisting of 8-membered-ring molecules. In experimental papers the melting temperature of sulfur is nearly always accompanied by the word "about". In our case the phrase "at about 120 °C" means that sulfur is definitely liquid at this temperature. However, sulfur can be melted at lower temperatures, but the process takes more time. We melted pure sulfur (99.999%, Alfa Aesar Co.) at 105° C, keeping it at this temperature in a sealed quartz cell for three days. The long times required for complete melting are typical for polymers, as for example for selenium [6], and they suggest the presence of polymeric molecules below the melting temperature. On the other hand, the observed interval for the melting temperature is consistent in part with results reported in Ref. 7, where it was ascribed to a difference in the melting temperature of different allotropes.

Close to 159 °C properties of sulfur change dramatically. Shear viscosity η, as measured using the Poiseuille technique, increases by four orders of magnitude over a temperature interval 25 C° [8]. More recent measurements performed with a sealed pyrex cell using a falling-ball technique indicate an even steeper change in η [9]. A specific feature resembling a λ-like singularity has been observed in temperature dependence of the heat capacity [10, 11]. Changes also occur in the temperature dependence of the density, ρ [12], and in the sound velocity, $c$ [13, 14]. The color of sulfur becomes darker, and the refractive index exhibits a minimum [15]. A minimum is also observed in the dielectric permittivity [16]. Electron spin resonance (ESR) spectra indicate a sharp increase of number of unpaired electrons. The ESR signal is interpreted as due to the appearance of linear polymers. An estimated degree of polymerization (average number of monomers in a polymeric chain) near the transition temperature $T_p$ is on the order of $10^6$ [17]. Finally, molten sulfur quenched from temperatures below 159°C is easily soluble in carbon disulfide, whereas sulfur quenched from higher temperatures yields an insoluble fiber-like residue, whose mass fraction Φ (also called extent of polymerization) grows with the quenching temperature and reaches saturation at about 300° C [2, 18]. The data for η, ESR spectra, and Φ suggest that liquid sulfur above 159°C is a solution of linear chain polymers in a monomeric solvent.

On the other hand, measurements of the neutron diffraction spectra [19] did not reveal significant changes in structure of sulfur molecules. This allowed authors to suggest that the transition of sulfur to the high viscous state is not due to polymerization but due to percolation of different molecular units with similar local configuration, but different thermal stabilities. Measurements of Raman spectra [20] likewise did not indicate an anomaly at the transition temperature. These Raman data can be viewed as supporting the percolation hypothesis of the aforementioned work. On the other hand, the results of both Raman and neutron diffraction spectra are also consistent with the standard view that the $S_8$ monomers convert to polymeric chains at $T \geq T_p$ and that the local structures of the polymers, such as bond lengths and bond angles, remain nearly the same as in the monomers [1]. In the most recent measurements of Raman spectra [21] a change in shape of a resonance peak was found, consistent with the polymerization hypothesis.

The properties of sulfur near the polymerization transition are completely reversible as is the case for a continuous phase transition. Wheeler et al. [22], Cordery [23], and Anisimov et al. [24] treat the polymerization in sulfur as a second order phase transition in a weak external field, using a Heisenberg n-vector model in the limit n → 0, the Ising model (the n-vector model with n=1), and Landau theory supplemented with fluctuation theory, respectively. The order parameters in these models are proportional to $\Phi$. It was shown [22], that an earlier theory by Tobolsky and Eiseberg [25], which successfully describes the temperature dependence of the degree and the extent of polymerization, can be considered a mean-field approximation for the second order phase transition. The second-order phase transition theories of the polymerization appear to be consistent with the experimental data on heat capacity. Agreement with other properties, such as density, looks less convincing [3]. These properties, however, do not exclude a possibility that a modified version of the second-order phase transition model could be applied to the equilibrium polymerization. It is worth noting that the polymers are linear chains in the models of Wheeler et al. [22] and Anisimov et al. [24], whereas they are loops in the Cordery's model [23].

Using the second order phase transition theory Anisimov et al. calculated sound velocity $c(T)$ for sulfur at the limit of zero frequency. A sharp minimum with a depth of about 10% from the value of $c$ just below $T_p$ has been predicted [24]. Taking into account that the minimum in $c$ is a general feature of the second order phase transitions [26], one expects that accurate measurements of acoustical properties will provide the ultimate check of applicability of this approach to the phenomenon of equilibrium polymerization.

Dudowicz, Freed and Douglas (DFD) proposed an alternative theory for the equilibrium polymerization [4], based on a mean-field Flory-Huggins-type incompressible lattice model. The lattice contains three kind molecular species: monomers, initiators, and chain polymers. This theory allows equilibrium properties, such as extent of polymerization and heat capacity, to be calculated; however compressibility and acoustical properties are out of scope of this model. The DFD theory agrees well with experimental data for α-methylsterene [3], which exhibits the equilibrium polymerization upon cooling. According to the DFD theory the polymerization is not a thermodynamic phase transition, but can mimic one in the limit of zero concentration of the initiator. In particular, the heat capacity approaches the λ-like shape in this limit. The DFD-model can be applicable to sulfur, for which the condition of the low initiator concentration is replaced by a condition of a small value of a constant of the initiator reaction (this constant for sulfur is on the order of $10^{-12}$ [25]). The principal feature distinguishing polymerization transition in the DFD-model from that in the second order phase transition models is that in the former $d\Phi/dT$ is a continuous function of $T$ near $T_p$, whereas in the latter models $\Phi(T)$ is zero at $T < T_p$ and has infinite slope at $T_p$.

Recently we measured the velocity and absorption of sound in liquid and supercooled sulfur at temperatures up to 200 °C and at frequencies from 5 to 22 MHz [14]. We observed no dispersion, but we did observe a pronounced change in the slope of the temperature dependence of the sound velocity near the polymerization transition, instead of the minimum following from the second order phase transition model. Assuming that there is no dispersion down to zero frequency, the extent of polymerization $\Phi(T)$ was calculated. The results are consistent with data on the extent of polymerization as measured under static conditions [18]. Surprisingly, we observed no change in the absorption of sound.

According to classical Navier-Stokes hydrodynamics [27], sound absorption α in liquids is described as

$$\alpha = [(4\eta/3 + \zeta) + \kappa(C_v^{-1} - C_p^{-1})]\omega^2/2\rho c^3 , \qquad (1)$$

where η and ζ are coefficients for the shear and bulk viscosities, respectively, κ is the coefficient of thermal conductivity, $C_v$ and $C_p$ are heat capacities, and $\omega = 2\pi f$ is the angular frequency. According to Eq. (1) the sound absorption near $T_p$ should increase by four orders of magnitude as the shear viscosity does [8]. For a second-order phase transition an additional increase should be expected due to increase of ζ [28].

Because we observed no such increase in absorption we tentatively proposed that the behavior might be associated with visoelasticity, which is typical for polymeric melts [29]. However, if this were the case, the visoelasticity could mask the minimum in $c$, and therefore the second order phase transition interpretation could be valid.

The goals of this work are the following.

1. Resolving the paradox about inconsistency of sound absorption data [14] with the classical hydrodynamics formula (1). If the source of the discrepancy is visoelasticity, then our assumption about absence of sound dispersion down to zero frequency is incorrect and therefore the interpretation of polymerization in sulfur as a second order phase transition cannot be ruled out. And vice versa: if melted sulfur is not viscoelastic liquid, then the polymerization transition is not a second-order phase transition, and the search for a comprehensive model of the equilibrium polymerization phenomenon (perhaps on the basis of the DFD model) should be continued. For this purpose extended acoustical measurements reported below, have been performed.

2. An additional check can be made on the basis of Raman spectrum measurements: if the polymerization transition were a second order phase transition, then the discontinuous change in $d\Phi/dT$ at $T_p$ should lead to a change in the intensities of the peaks going through the transition. The Raman measurements reported below were designed to test this possibility.

3. Finally, the ESR measurements have been performed to verify the type of the polymers in liquid sulfur: are they loop-like species, as proposed on bases of the data on neutron diffraction [19], or the polymers are the long liner chains. In addition we intended to check the presence of polymers in liquid sulfur in low-viscous state above and below the melting temperature.

The next section briefly describes the experimental techniques. Results of the measurements are presented in Section 3. These results are discussed in Section 4. Section 5 contains a summary.

2. EXPERIMENTAL

Because the properties of sulfur are strongly affected by impurities and dissolved gases [1-3], careful precautions were taken to preserve the purity of the sulfur and to isolate it from air. We used sulfur of 99.9999% purity from Comica Ltd, Canada and of 99.999% purity from Alfa Aesar Co. Sealed quartz and pyrex cells were employed for the Raman and ESR measurements, respectively. These cells were filled with sulfur via sublimation under vacuum as described previously [14]. Sulfur that was carefully degassed and sealed in a quartz cell was used for the low-frequency sound velocity measurements. The sound absorption measurements utilized a cell made out of stainless steel, which was kept under vacuum. The measurements of sound impedance were made with a quartz cell open to the air because of the need to access the end of the rod to remove a residual drop of sulfur. The consistency of the data obtained in the different cells suggests that any effects of contamination are less than the experimental uncertainties.

2.1. Electron spin resonance.

ESR spectra of sulfur have been measured from room temperature to 200 °C using a Bruker Instruments ESR spectrometer operating at a microwave frequency of 9.5 GHz. The spectra were taken with a modulation amplitude of 1 mT using standard first-harmonic detection techniques. The sample was placed in a quartz, flow-through dewar and heated by passing hot nitrogen gas through the dewar. The temperature was calibrated outside of the microwave cavity using a thermocouple in the sample tube. The error in the temperature measurements was ± 1 K.

2.2. Raman Spectroscopy.

Raman spectra were obtained from solid sulfur at room temperature and from liquid sulfur at temperatures near the polymerization transition (from 150 to 170 °C). The spectra were excited with the 488 nm and 514 nm lines of an argon ion laser, dispersed with a 0.75 m triple spectrometer, and detected with a liquid-nitrogen-cooled charge-coupled-device (CCD) camera. The sample was held in an evacuated quartz ampoule in a heated copper block with a small opening for optical access. The temperature was monitored with three standard copper-constantan thermocouples, one at the level of the opening, one above and one below, to monitor the temperature over the entire sample volume. The temperature of the sample was measured to within a tenth of a degree for each measurement.

2.3. Acoustics.

To check the possible dispersion, we extended the frequency range of the sound velocity measurements down to 500 kHz, which is a factor of ten lower than our previous measurements [14]. We also performed direct measurements of sound absorption by varying the length of the sample. A final check of the viscoelastic hypothesis was made by measuring sound impedance for both longitudinal and transverse sound waves. All the experiments were performed under steady state conditions using a pulsed phase-sensitive technique described elsewhere [30].

The theoretically predicted minimum in the sound velocity at the polymerization transition [24] is related to thermodynamic or "zero frequency" sound, which demands that the frequency be less than the lowest relaxation frequency of the system. The relaxation frequency $f_r$ can be estimated as $f_r \equiv (\tau_r)^{-1} \sim (\eta \cdot \beta)^{-1}$, where $\beta$ is compressibility of fluid [31]. For simple liquids $f_r$ is on the order of 10 to 100 GHz, and therefore sound velocity at frequencies in the MHz range is essentially the same as that for zero frequency. In polymeric systems the situation is different. Configuration relaxation associated with collective displacements of big groups of atoms in macromolecules produces a distribution of relaxation times [32], the longest of which, $\tau^*$, can be in the range of microseconds to even milliseconds. If the time scale of the distortion is less than $\tau^*$ the system possesses an effective Young's modulus and behaves like an elastic solid; if the time scale of the distortion is greater than $\tau^*$ the system responds to shear stress like a viscous liquid. This is the essence of the viscoelastic effect.

2.3.1. Low frequency sound velocity.

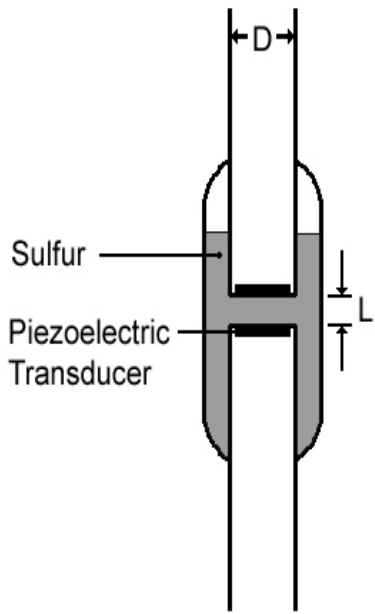

Fig. 1. Schematic diagram of the quartz cell for measurements of sound velocity.

The cell for the low frequency measurements is shown in Fig. 1. The cell was made out of quartz and has two parallel windows. To fulfill the plane-wave approximation, the diameter D was chosen 25 mm, which is for ten times greater than the sound wavelength $\lambda$. The sample length L, about 13 mm, was calibrated using water.

High purity sulfur powder was loaded in the cell, melted, and degassed under a vacuum of $10^{-6}$ mm Hg, after which the cell was sealed. Experiments with sulfur in quartz cells require care because solid sulfur must be reheated very uniformly to avoid small temperature gradients that will crack the cell due to expansion on local melting. The cell was held in a glass oil bath as described in Ref. 14. The temperature of the sample was measured using three cooper/constantan thermocouples and controlled by an oil thermostat (6330, Hard Scientific, Inc) supplied with a home made pump. The temperature variation over the sample was less than 2 °C. High-temperature ceramic piezoelectric transducers (Staveley Sensors Inc.) with a principal frequency of 500 kHz were attached to the windows via non-polymerized epoxy resin. The total fractional uncertainty of the sound velocity measurements was 0.2%.

2.3.2. Sound absorption.

Previous direct measurements of the sound absorption coefficient α in liquid sulfur [33] found a small increase in α near the polymerization transition. These measurements were performed at frequencies from 6 to 15 MHz using an open chrome-plated brass cell with a moving aluminum reflector. The authors noted instabilities in the reflected signal caused by forming and collapsing of gas bubbles in the liquid sulfur. In the present measurements this effect was excluded by keeping sample under vacuum (~ $10^{-4}$ mm Hg) in a hermetically sealed cell, shown schematically in Fig. 2. The cell was inserted into a thick-walled copper block to produce a uniform temperature over the sample. The temperature was controlled with a Smart-3 temperature controller, phase angle power controllers PC 120/240-PA (both of Temp, Inc), and three copper-constantan thermocouples. For longitudinal sound measurements, single crystal lithium niobate transducers (12 mm diameter) with a principal frequency of 7 MHz were attached to the movable quartz rod and to the bottom of the cell as shown in Fig. 2. The lower transducer served as the transmitter, and the upper one as the receiver. The rod was displaced using a coaxial screw (not shown). The maximum displacement of the rod was 25 mm. This displacement was measured using a Cen-Tech. Co indicator with a precision of 0.002 mm. The amplitude of the signal was measured on an oscilloscope screen with an uncertainty of about 5%. The total uncertainty for the sound absorption data was 10%.

To decrease potential contamination of the sulfur by a reaction with stainless steel, we minimized the time of the measurements. Two runs were performed, each one with a new load of sulfur. The first run took about 2 hours and the second run over more than 6 hours. The data were identical in both runs indicating that contamination, if any, does not affect the measurements within the error. After each run the cell was washed with carbon disulfide, and no traces of corrosion were noticed.

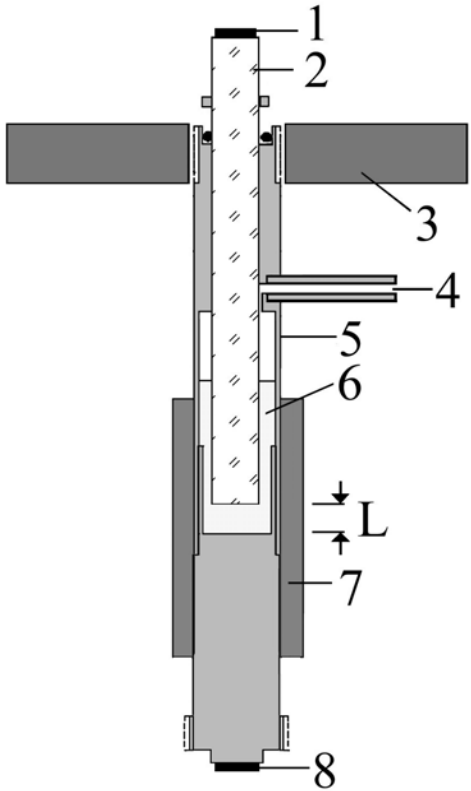

Fig. 2. The cell used for measurements of sound absorption.

1, piezoelectric transducer (receiver); 2, movable quartz rod; 3, platform; 4, pipe to vacuum pump; 5, cell body made of stainless steel; 6, sulfur; 7, cooper block; 8, piezoelectric transducer (transmitter).
L is the changeable sample length.

2.3.3. Sound impedance.

If sulfur is viscoelastic it should support propagation of transverse sound waves [34]. We first attempted to perform direct measurements of the speed of transverse sound using a quartz cell as described in Ref. 14. The sample length was 0.8 mm. Quartz shear-mode transducers with a principal frequency of 5 MHz were attached to the buffer rods via hardened epoxy resin. Measurements were performed at 5, 15, and 25 MHz at temperatures up to 180 °C. At all three frequencies the transmitted signal disappeared on melting and reappeared on solidification; no transmitted signal was observed either below or above the polymerization transition temperature.

The absence of a transmitted signal may result either from the absence of the viscoelastic effect or from strong absorption of the shear waves in polymerized sulfur. To check these two versions, we measured the acoustical impedance Z of the sample. In this experiment, an rf pulse of sound is sent down a quartz rod, and, the amplitude and the phase of the reflected (from the end of the rod) pulse are measured. The measurement is performed twice, once when the rod is in contact with the liquid and once when it is not. The ratio of the amplitudes of the reflected signals $r$ and the phase shift between them $\theta$ are associated with sound velocity $c$ and sound absorption $\alpha$ as follows [35]:

$$r = \frac{Z_q - Z'}{Z_q + Z'} = \frac{Z_q - \rho c}{Z_q + \rho c}, \tag{2}$$

and

$$\sin \theta = \frac{Z''}{Z_q} \cdot \frac{(1 + r)^2}{2r} = \alpha \rho c \lambda \frac{(1 + r)^2}{4\pi r Z_q}, \tag{3}$$

where $Z'$ and $Z''$ are the real and imaginary parts of the sound impedance of the fluid, $Z_q$ is the sound impedance of quartz ($Z_q$ is assumed real). Details of the experimental set-up and measuring procedure are published elsewhere [36].

3. RESULTS

The ESR spectra are shown in Fig.3 for several temperatures. The amplitude of the signal, and hence the number of chains, increases with temperature above 166 °C. Below this temperature no spectrum was observed, indicating a spin density below the detection limit of the spectrometer, which is on the order of $10^{15}$ spins/cm$^3$.

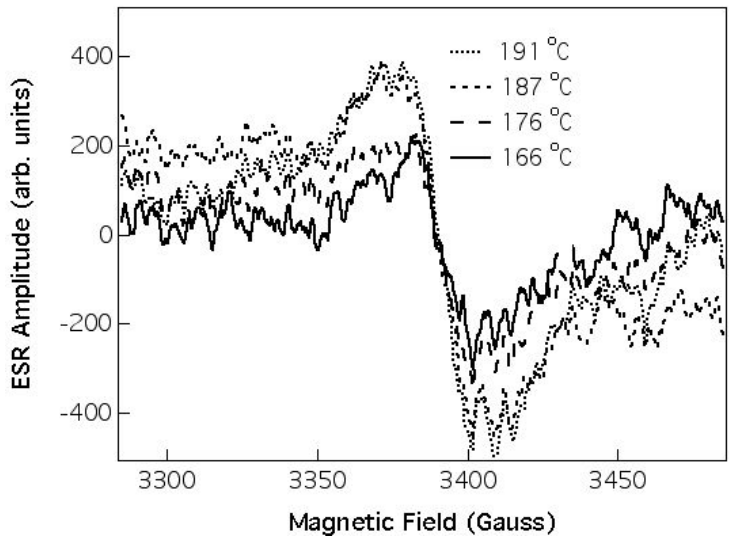

Figure 3. Electron spin resonance spectra at four temperatures above the transition temperature.

Superimposed Raman spectra at different temperatures are shown in Fig. 4. No dramatic change in spectral features was observed. For an $S_8$ molecule, the 475 cm$^{-1}$ peak is the unresolved combination of the $a_1$ and $e_1$ Raman-active modes, while the 218 cm$^{-1}$ peak is due to an $a_1$ mode [21, 37]. Over the temperature range from 150 to 170 °C there is a monotonic decrease in the intensities of the peaks with increasing temperature.

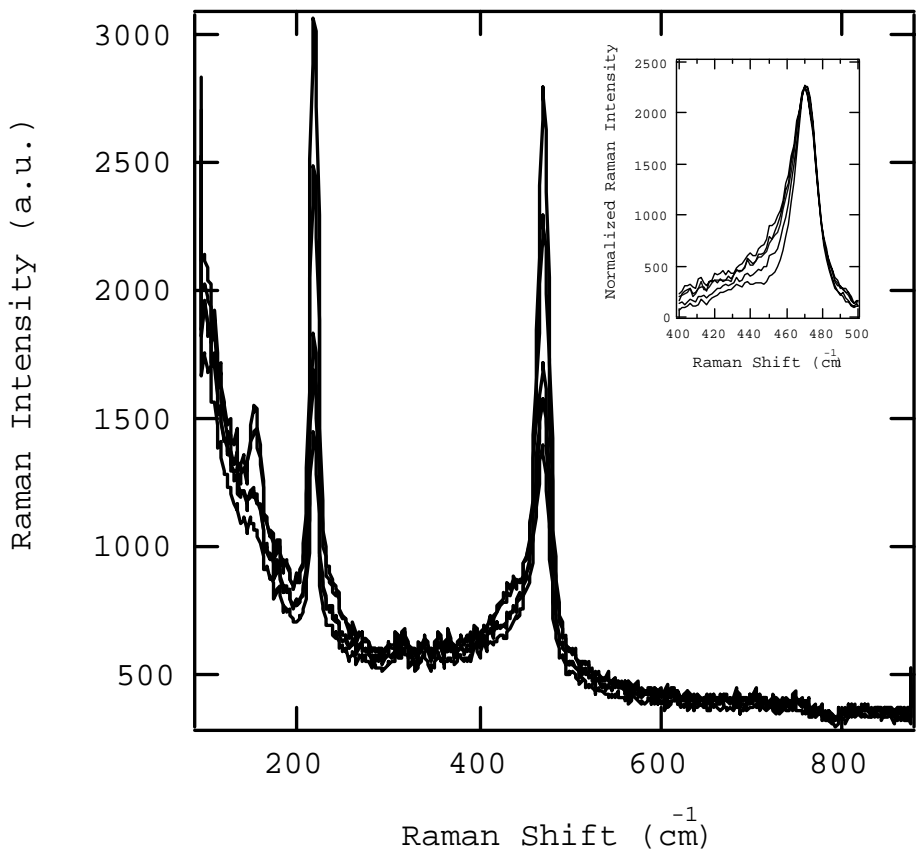

Fig. 4. Temperature dependence of Raman spectra from 150 to 170 °C. The inset shows the spectra near 475 cm$^{-1}$ normalized to the same peak height. The relative increase of the low-frequency component in the inset is indicative of an increase in polymerization [21] as described in the text.

In Figure 5, the intensity of the peak at 475 cm$^{-1}$ is plotted as a function of temperature. The solid and open triangles indicate spectra taken while the sample was heating and cooling, respectively. The intensity varies smoothly through the polymerization transition. This temperature dependence could, in principle, be due to changes in the Raman cross-section [38] with temperature, but most likely it is due to the temperature dependence of the optical absorption coefficient [15]. At the energy of the exciting laser, the absorption coefficient increases with increasing temperature effectively decreasing the active volume of the sample.

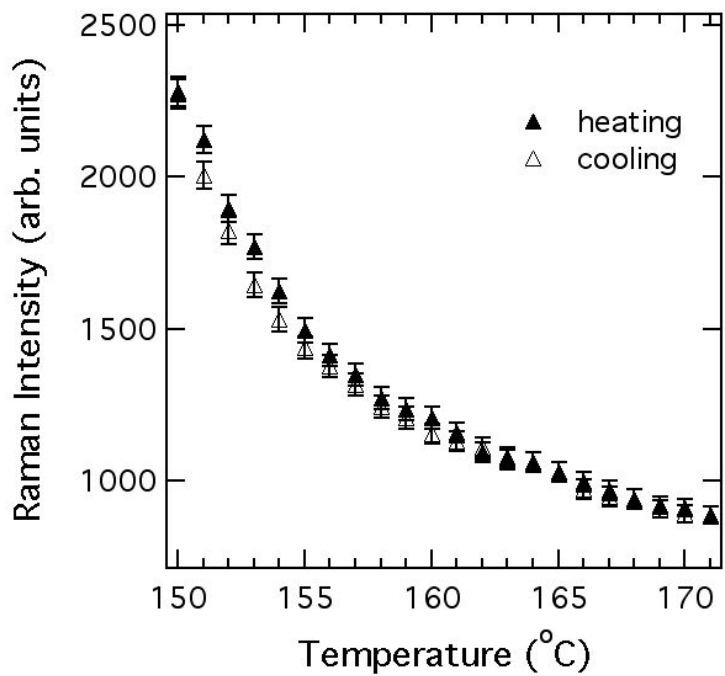

Figure 5. Intensity of the Raman peak at 475 cm$^{-1}$ near the polymerization transition temperature.

Recently Kalampounias et al. [21] analyzed in detail the shape of the peak near 475 cm$^{-1}$. Their analysis shows that the unresolved combination of Raman scattering is due to ring and chain components of the liquid, which may be fit to a sum of Lorentzian peaks to predict the extent of polymerization in the liquid. A similar analysis of our data is shown as an inset in Figure 4. The shoulder near 430 cm$^{-1}$ is shown to increase relative to the overall peak intensity at temperatures above the polymerization transition indicating an increase of the polymers in agreement with Ref. 21.

Results of measurements of the sound velocity at 500 kHz are shown in Fig.6 along with the data from previous measurements [14] (the new data are shown with larger symbols). There is no dispersion in the sound velocity down to 500 kHz. The time required to take each experimental point varied from 5 to 30 minutes. In contrast to the kinetic effect that occurs at melting, no time dependence or hysteresis of the acoustical data were observed in liquid sulfur. Both the phase and the amplitude of the transmitted signal followed the temperature of the sample; as soon as the temperature was stabilized, the signal was stable as well.

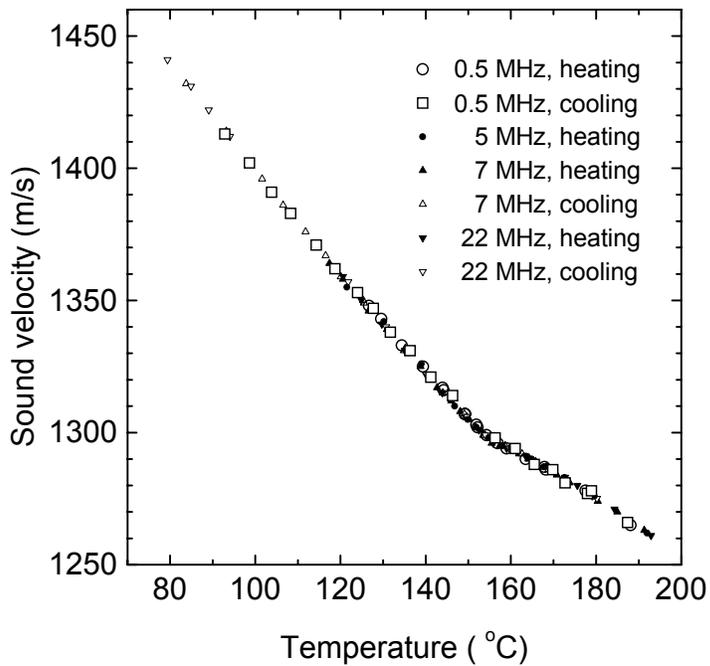

Fig.6. Sound velocity in liquid and supecooled sulfur.

To check for a dependence of the experimental data on the manufacturer of the sulfur, we repeated the experiment at 7 MHz with the cell described in Ref. 14 using sulfur received from S. C. Greer (99.9998% purity, Gallard-Schlesinger Chemical Corp.), which was identical to that used for density measurements [12]. The results, which are omitted from Fig. 6 for simplicity, were identical to those reported in Ref.14 and shown in Fig. 6.

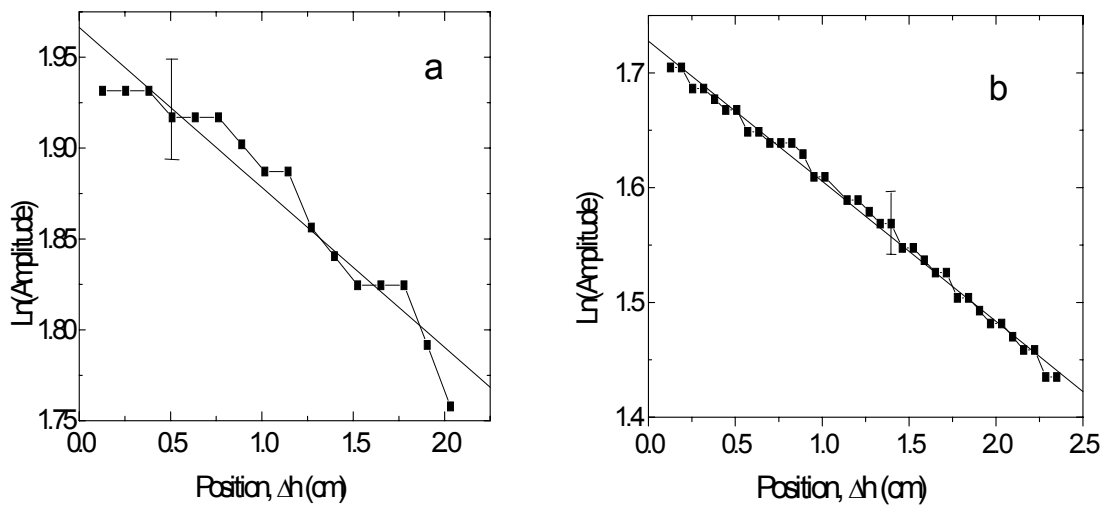

Fig.7. Amplitude of the transmitted ultrasonic signal as a function of sample length (position of the upper buffer rod). a, T=120 °C; b, T=170 °C. See text for details.

Data for sound absorption at two representative temperatures are shown in Fig.7. The absorption coefficient α in nepers per centimeter was calculated by fitting the data at each temperature to an exponential curve. Results of these fits at all temperatures are shown in Fig. 8. There is no specific feature in the sound absorption at the polymerization transition in agreement with previous measurements [14].

The data on sound velocity, and absorption still could be, in principle, consistent with the viscoelastic interpretation, if the relaxation frequency $f^* \equiv 1/\tau^*$ is below 100 kHz. For an ultimate check of the viscoelastic hypothesis, we performed measurements of the shear acoustical impedance.

The apparatus was first tested with longitudinal sound. The amplitude ratio, $r$, and the phase shift θ of the longitudinal acoustical impedance at 7 MHz are shown in Fig. 9. The data for $r$ agree well with those calculated from precision density [12] and sound velocity [14] measurements. The measured phase shift θ is essentially zero over the entire temperature range. Using only the shear viscosity term of Eq. (1) and viscosity data [8] one can estimate θ to be on of the order of $10^{-4}$ for low temperatures and close to π/2 above the polymerization transition. Clearly, this estimate disagrees with the experimental data for θ. However, the data are consistent with the sound absorption measurements of this work as well as with the data reported in Refs. 14 and 33.

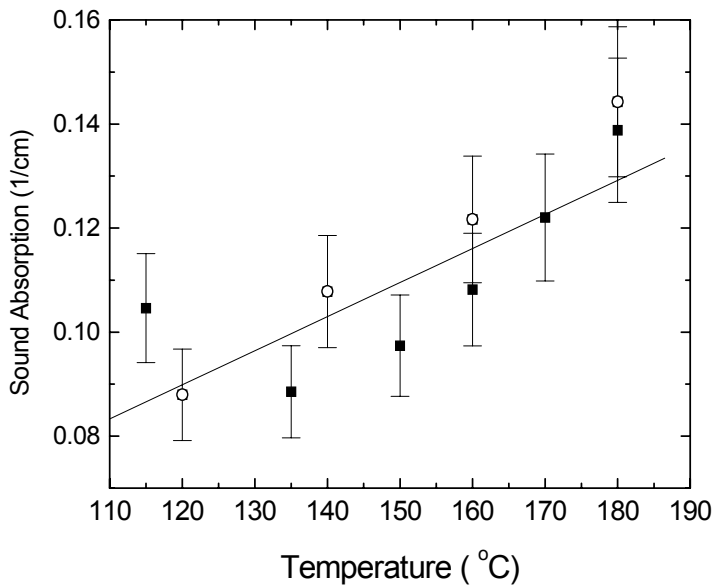

Fig.8. Absorption of sound in liquid sulfur at 7 MHz. Open cycles denote points taken on the first run, and solid squares represent points taken on the second run. See text for details.

Measurements of the shear wave impedance at 5, 15, and 25 MHz and at temperatures up to 180 °C yielded no change in either the amplitude ($r = 1$) or the phase ($\theta = 0$) of the reflected signal. The same result ($r \neq 1$ for longitudinal waves, and $r = 1$ for transverse waves) occurs in water, which is definitely not viscoelastic.

## 4. DISCUSSION

The density of spins above the polymerization transition temperature, as estimated from the ESR, is shown in Fig.10. Although the absolute densities are known only within about a factor of three, the relative densities shown in Fig. 10 are accurate within the errors shown. The absolute densities are derived by comparison to a standard ESR sample. Below 160 °C the density of unpaired spins is less than the detection limit of the spectrometer.

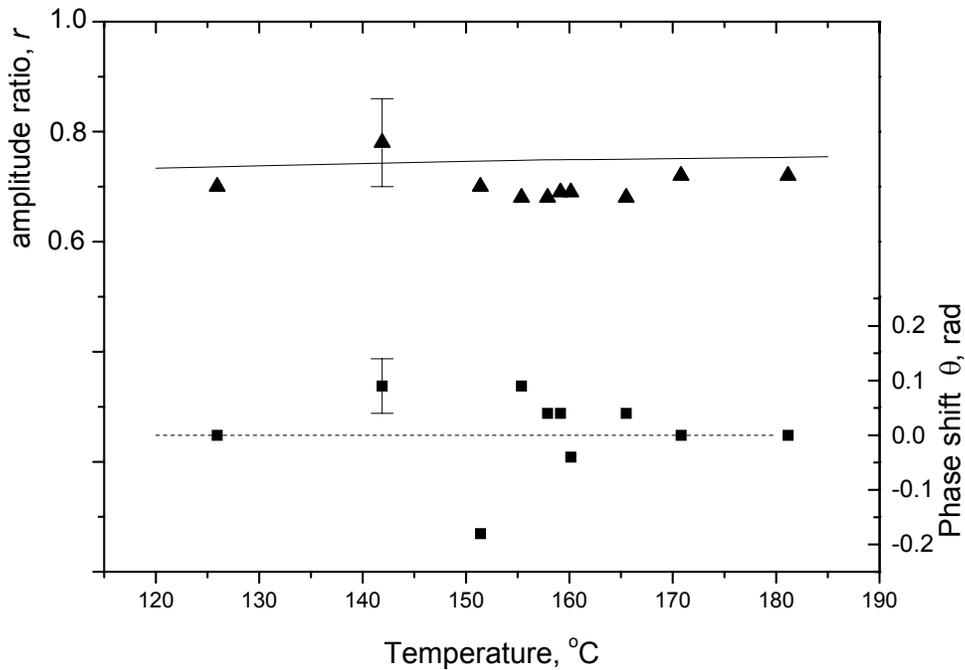

Fig. 9. Amplitude ratio $r$ (triangles) and phase shift $\theta$ (squares) measured with longitudinal waves at 7 MHz. The solid line represents $r$ calculated from Eq. (2) using precision data on density [12], sound velocity [14], and sound impedance of quartz [39]. The dashed line represents the phase shift calculated using Eq. (3) and the sound absorption data measured in this work.

The ESR signal comes from unpaired electrons that occur at the ends of the chain molecules [40]. Using the ESR data in combination with the data on the extent of polymerization $\Phi$ derived from the sound velocity data [14], the degree of polymerization $D$ (the average number of monomers in a polymer or the average polymer length) was estimated as follows:

$$D \sim \rho\, N_{av} \Phi / (n\, \mu), \qquad (4)$$

where $N_{av}$ is Avogadro's number, $n$ is the number of spins per unit volume, and $\mu$ is the atomic mass. Using the spin densities from Fig.10 and the extent of polymerization calculated from the sound velocity [14], the average chain length near the polymerization temperature is on the order of $10^5$ monomers, which is consistent with the results of a more detailed ESR study [17].

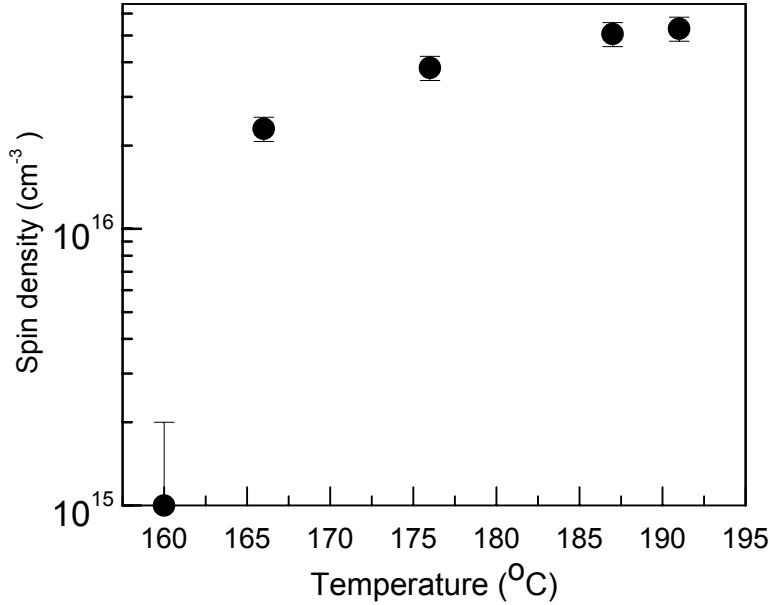

Fig. 10. The density of unpaired spins as determined by ESR. An upper bound for the number of spins below the transition temperature (160 °C) is at most $10^{15}$ cm$^{-3}$.

The ESR measurements also allow us to estimate an upper limit for the possible number density of the chain molecules and their degree of polymerization. Assuming that $n$ is of the order of the detection limit and $\Phi$ is of the order of the error in measurements of the extent of polymerization ($\approx 1$ %), one obtains that $D$ at low temperatures might be as big as $10^5$. That is, the polymer length at low temperatures might be of the same order of magnitude as the length above the polymerization temperature. It is possible that existence of even small amount ($\leq 1$ %) of such macromolecules in solid sulfur might affect kinetics of melting.

The relative independence of the Raman spectra on temperature through the polymerization transition suggests that there is no visible difference in interatomic bonding between polymers and monomers that would affect the vibrational spectra. The smooth behavior of the Raman intensity is consistent with the DFD model.

The null results from the measurements of transverse sound velocity and impedance, and the data on sound velocity at 500 kHz confirm that there is no dispersion down to zero frequency, and therefore acoustical properties measured in the MHz range can be treated as the data in the limit of zero frequency, as previously suggested [14]. This conclusion is consistent with Brillouin scattering spectra for liquid sulfur, where only the longitudinal mode peak was observed [41]. Contrarily, the Brillouin spectrum for viscoelastic liquid boron oxide ($B_2O_3$) exhibits well-defined peaks for both longitudinal and shear modes [42]. Thus we conclude that melted sulfur is not a viscoelastic liquid, and therefore the polymerization phenomenon is not the second order phase transition. This conclusion is consistent with predictions of the DFD-model [4]. Apart from that the data on $\Phi(T)$ [2, 3, 14, 18, 21] in the vicinity of the polymerization temperature $T_p$ appear to be more consistent with the DFD-model, than with the second order phase transition models.

In addition, the absorption of sound in polymeric sulfur is not proportional to the shear viscosity, as would be predicted from Eq. (1). One may speculate that, in a solution of monomers, initiators, and polymers, the low impedance of the monomeric component shunts the impedance of sound from the polymers. Therefore, the overall impedance of the system is determined by the component of lowest viscosity.

CONCLUSIONS.

The polymerization transition in liquid sulfur is a process of building long chain polymers from ring-like monomer units. The Raman spectra show that the interatomic bonding in sulfur remains essentially unchanged in both the monomeric and polymeric states. Acoustical properties of sulfur measured in the MHz frequency range, represent these properties in the thermodynamic limit. Therefore, sound velocity data can be used to characterize the mass fraction of polymers above the polymerization transition, as was suggested in Ref. 14. The equilibrium polymerization in sulfur is not a second order thermodynamic phase transition but rather an equilibrium mixture of monomers and polymers, as described by the DFD-model [4].

Finally, the absence of anomaly in sound absorption near the polymerization transition testifies that eq. (1) is not applicable for the polymeric solution.

ACKNOWLEDGEMENTS

One of us (VK) gratefully acknowledges Anneke Levelt Sengers for turning his attention to phenomena in liquid sulfur, and he is deeply grateful to Michael E. Fisher, Sandra C. Greer and Mikhail A. Anisimov for interest and encouragement. Grateful thanks are due to P. Craig Taylor for general support and valuable comments, and to Janica L. Whitaker for the help in the ESR measurements. The authors acknowledge the donors of the Petroleum Research Fund, administrated by the American Chemical Society (under grant # 326802-AC5), and the National Science Foundation (under grant # DMR-0073004) for support of this research.  CEI and CLMcD gratefully acknowledge the summer fellowship program of the Petroleum Research Fund.

References.
1. B. Mayer, Chem. Rev. **76**, 367 (1976)
2. R. Steudel, Topics in Current Chemistry **230,** 81 (2003)
3. S. C, Greer, Annu. Rev. Phys. Chem. **53**, 173 (2002); J. Phys. Chem. B **102**, 5413 (1998).
4. J. Dudowicz, K. F. Freed, and J. F, Douglas, J. Chem. Phys. **111**, 7116 (1999); **12**, 1002 (2000); **113**, 434 (2000).
5. See, for example, S. Harrington, R. Zhang, P. H. Poole, F. Sciotino, and H. E. Stanley, Phys. Rev. Lett. **78**, 2409 (1997); K. Ito, C. T. Moynihan and C. A. Angell, Nature **398**, 492 (1999); T. Morishita, Phys. Rev. Lett. **87**, 105701 (2001); P.G. Debenedetti and H.E. Stanley, Physics Today, June 2003, p. 40
6. W. B. Payne, V. F. Kozhevnikov and P. C. Taylor, to be published
7. M. Thackray, J. Chem. Eng. Data **15**, 495 (1970)
8. R. F. Bacon and R. Fanelli, J. Am. Chem. Soc. **65**, 639 (1943)
9. Ruiz-Garcia, E. M. Anderson, and S. C. Greer, J. Phys. Chem. **93**, 6980-6983 (1989).
10. F. Feher and E. Hellwig, Z. Anorg. Allg. Chem. **294**, 63 (1958).
11. E. D. West, J. Am. Chem. Soc. **81**, 29 (1959).
12. K. M. Zheng and S. C. Greer, J. Chem Phys. **96**, 2175 (1992).

13. D. L. Timrot, M. A. Serednitskaya, and T. D. Chkhikvadze, Sov. Phys. Dokl. **29**, 961-963 (1984).
14. V. F. Kozhevnikov, J. M. Viner, and P. C. Taylor, Phys. Rev. B **64**, 214109 (2001).
15. K. Tamura, H. P. Seyer, and F. Hensel, Ber. Bunsenges. Phys. Chem. **90**, 581 (1986).
16. S. C. Greer, J.Chem. Phys. **84**, 6984 (1986).
17. D. C. Koningsberger, "The polymerization in liquid sulfur and selenium. An ESR study", Thesis, Technishe Hogeschool te Eindhoven, Eindhoven, 1971.
18. J. C. Koh and W. Klement, Jr. J. Phys. Chem. **74**, 4280 (1970).
19. R. Winter, C. Szornel, W-C Pilgrim, W.S. Howells, P.A. Egelstaff, J. Phys.: Condens. Matter **2**, 8427 (1990); Chr. Biermann, R. Winter, Chr. Benmore, P. A. Egelstaff, J. Non-Crystalline Solids **232-234**, 309 (1998).
20. A. T. Ward, J. Phys. Chem. **72**, 4133 (1968)
21. A. G. Kalampounias, K. S. Andikopoulos, S. N. Yannopoulos, J. Chem. Phys **118**, 8460 (2003).
22. J. C. Wheeler, S. J. Kennedy, and P. Pfeuty, Phys. Rev. Lett. **45**, 1748-1752 (1980); J. C. Wheeler and P. Pfeuty, Phys. Rev. A **24**, 1050 (1981)
23. R. Cordery, Phys. Rev. Letters **47**, 457 (1981).
24. M. A. Anisimov, K. I. Kugel, and T. Yu. Lisovskaya, High Temp. **25**, 165-173 (1987)
25. A. V. Tobolsky and A. Eisenberg, J. Am. Chem. Soc. **81**, 780 (1959).
26. H.E. Stanley, Introduction to Phase Transitions and Critical Phenomena (Clarendon Press, Oxford, 1971)
27. L. D. Landau and E. M. Lifshitz, *Fluid Mechanics* (Pergamon, Oxford, 1966).
28. M. A. Anisimov, *Critical Phenomena in Liquids and Liquid Crystals* (Gordon and Breach, Philadelphia, 1991).
29. P.-G. de Geness, Scaling Concept in Polymer Physics (Cornell University Press, Ithaca, 1979)
30. V. F. Kozhevnikov, D. I. Arnold, M. E. Briggs, S. P. Naurzakov, J. M. Viner, and P. C. Taylor, J. Acoustic Soc. Am. **106**, 3424 (1999).
31. J. C. Maxwell, Phil. Trans. Roy. Soc. **157**, 49 (1867).
32. A. Yu. Grossberg and A.R. Khokhlov, Statistical Physics of Macromoecules (AIP Press, New York, 1994).

33. A. W. Pryor and E. G. Richardson, J. Phys. Chem. **59**, 14 (1955).

34. W. P. Mason, W. O. Baker, H. J. McScimin, and J. H. Heiss, Phys. Rev. **75**, 936 (1949).

35. H. J. McScimin, in Physical Acoustics, Edited by W. P. Mason, v. 1, p. 271 (Academic Press, N.Y., 1964).

36. J. K. Olson, W. B. Payne, C. McDonald, C. Inglefield, V. F. Kozhevnikov, and P. C. Taylor, International Journal of Thermophysics, submitted.

37. D. W. Scott, J. P. McCullough, and F. H. Kruse, Journal of Molecular Spectroscopy **13**, 313 (1964).

38. D. J. Gardiner, *Practical Raman Sprectroscopy* (Springer-Verlag, NY, 1989)

39. J. Handbook of Physical Quantities, edited by L. G. Grigoriev and E. Z. Meilikhov (CRC, New York, 1997)

40. D. M. Gardner and G. K. Fraenkel, J. Am. Chem. Soc. **78**, 3279 (1956).

41. A. D. Alvarenga, M. Grimsditch, S. Susan, and S. C. Rowland, J. Phys. Chem. **100**, 11456 (1996).

42. M. Grimsditch, R. Bhadra, L. M. Torell, Phys. Rev. Lett **62**, 2616 (1989).